\documentstyle[aps,pre,preprint]{revtex}
\begin{document}
\draft
\tightenlines
\title{Hysteresis in vibrated granular media}
\author{A.\ Prados, J.\ Javier Brey}
\address{F\'{\i}sica Te\'orica, Facultad de F\'{\i}sica, Universidad
         de Sevilla, Apdo.\ de Correos 1065, E-41080 Sevilla, Spain}

\author{B.\ S\'anchez-Rey}
\address{F\'{\i}sica Aplicada, E.\ U.\  Polit\'ecnica, Universidad de Sevilla,
        Virgen de \'Africa 7, E-41011  Sevilla, Spain}
\date{today}
\maketitle
\begin{abstract}
Some general dynamical properties of models for compaction of granular media
based on master equations are analyzed. In particular, a one-dimensional lattice
model with short-ranged dynamical constraints is considered. The stationary
state is consistent with Edward's theory of powders. The system is submitted to
processes in which the tapping strength is monotonically increased  and
decreased. In such processes the behavior of the model resembles the
reversible-irreversible branches which have been recently observed in
experiments. This behavior is understood in terms of the general dynamical
properties of the model, and  related to the hysteresis cycles exhibited by
structural glasses  in thermal cycles. The existence of a ``normal'' solution,
i.\ e., a special solution of the master equation which is monotonically
approached by all the other solutions, plays a fundamental role in the
understanding of  the hysteresis effects.

\end{abstract}

\section{Introduction}
\label{sec1}

The phenomenology of granular materials is very rich. One of the most
characteristic dynamical behaviors is compaction. If a system of grains in a
loosely packed configuration is submitted to vertical vibrations, or tappings,
it approaches slowly a more compact state. The relaxation is well described by
an inverse logarithmic law, and both the relaxation function and the density of
the ``stationary'' state seem to depend only on a dimensionless parameter
characterizing the vibration intensity \cite{KFLJyN95}.

More recently, it has also been shown that granular materials exhibit
irreversible-reversible cycles if the vibration strength is increased and
decreased alternately. When the system is tapped with increasing intensity
starting from a loosely packed state, the density shows a non-monotonic
behavior, with a maximum at a certain value of the intensity of vibration. If,
afterwards, the granular medium is tapped  with decreasing intensity, the
density increases monotonically, and hysteresis effects show up. Interestingly,
if the vibration intensity is again increased, the density follows a curve that
is approximately equal to the evolution in the decreasing intensity process,
being thus ``reversible'' \cite{NKBJyN98,Ja98}. Of course, the rate of variation
of the vibration intensity is the same for all the processes described above.

The static properties of granular materials have been studied by Edwards and
co-workers \cite{EyO89,MyE89}. The starting point is the plausible idea that all
the ``microscopic'' configurations of a powder having the same volume are
equiprobable, provided that the powder has been prepared by extensive
manipulation, i.\ e., by processes which do not act on individual grains. This
has been called the ``ergodic hypothesis'' of powders. On this basis, it is
possible to make an analogy between the variables of a molecular  system and the
parameters characterizing the state of a powder. The volume of a powder is
analogous to the energy, while the entropy remains the same quantity, measuring
the available number of configurations. The derivative of the volume with
respect to the entropy is called the ``compactivity'' of the powder, and plays
the role of the temperature. The lowest compactivity corresponds to the densest
state (minimum volume), and the highest compactivity to the fluffiest stable
configuration (maximum volume). The stationary value of the volume is an
increasing function of the compactivity, like the energy increases with the
temperature in a thermal system.

It seems interesting to try to establish a connection between the dynamical and
the equilibrium properties of powders. It looks  reasonable that, if the steady
state is reached in a tapping process, there should be a relationship between
the vibration intensity and the compactivity of a powder. The tapping process
allows the system to explore the phase space of available configurations, and
the same role is played by the temperature  in  a thermal system. In this way,
it is tempting to explain the tendency of a granular system to move over an
almost reversible curve, when the tapping strength varies in time, as the
approach towards the stationary curve of the powder. If the above  is true, the
compactivity will be an increasing function of the vibration intensity,  because
the density over the ``reversible'' curve decreases with the tapping strength.
In fact, this kind of behavior has been found in a simple model of a granular
system described in terms of a Fokker-Planck equation \cite{EyG98}.

Also, glass formers exhibit hysteresis effects when cooled and heated through
the glass transition region. When a glass former is cooled, the system follows
the equilibrium curve until a certain temperature $T_g$, at which it gets frozen
due to the fast increase of the relaxation time. If this structural glass is
reheated from its frozen state, the equilibrium curve is only approached for
temperatures larger than $T_g$. This is due to the fact that the system starts
from a configuration in which the structural rearrangements are very difficult
\cite{Sch86}. In this way, also hysteresis effects show up in glass formers,
when they are submitted to thermal cycles.

The study of simple models have been very useful in order to understand, at
least qualitatively, the behavior of structural glasses. In particular,
hysteresis cycles are also shown by simple models when cooled and heated
\cite{MJPhD92,FyR95}. The analytical approach to thermal cycles is a very
difficult task, because it is necessary to solve the kinetic equations of the
model with time-dependent coefficients. Nevertheless, some models whose dynamics
is formulated by means of a simple master equation can be exactly solved
\cite{ByP94,PPhD93}. In those situations, the role played by the ``normal
solution'' of the master equation, which is monotonically approached by all the
other solutions \cite{ByP93}, has been shown to play a fundamental role.

Here we will consider a simple model for granular media, which has been
previously introduced \cite{BPyS99}. Its dynamical evolution is governed by a
master equation, with the transition rates given as functions of the tapping
strength. When the system is vibrated with a given intensity from a low density
state, the density of the system increases monotonically until reaching an
stationary value. The relaxation of the density is very slow, being very well
fitted by an empirical inverse-logarithmic law. Interestingly, the stationary
state of the system is described by Edward's theory of powders. Then, it seems
worth studying its behavior when the tapping strength changes in time for
several reasons. Firstly, in order to verify if its behavior resembles that of
real granular systems, showing the irreversible and reversible branches found in
experiments \cite{NKBJyN98}. Secondly, to understand whether the hysteresis
effects are related to the existence of a ``normal solution'' of the master
equation with time-dependent tapping strength.

The paper is organized as follows. In Sec.\ \ref{sec2} some general properties
of models for granular media based on a master equation formulation of the
dynamics are considered. The existence of a ``normal solution'' of the master
equation is analyzed for the case of time-dependent transition rates. Also, the
conditions to be verified by the law of variation of the tapping strength in
order to guarantee that the system approaches the steady state curve are
discussed. Section \ref{seca} is devoted to the analysis of the normal solution
by means of Hilbert's expansion of the master equation. A quite general
expression for the first order deviation of the normal solution from the
stationary curve is obtained. The specific model to be considered is presented
is Sec.\ \ref{sec3}, as well as a brief discussion of its equilibrium state.
Section \ref{sec4} deals with the behavior of the model in processes with
time-dependent vibration intensity. Firstly, processes in which the tapping
strength decreases in time are studied. The existence of a phenomenon similar to
the laboratory glass transition is analyzed. Secondly, we discuss processes with
increasing vibration intensity. The role played by the normal solution turns out
to be fundamental. Finally, the main conclusions of the paper are summarized in
Sec.\ \ref{sec5}.

\section{Some general dynamical properties}
\label{sec2}

Let us consider a model system whose dynamics is described by means of the
master equation \cite{vk92}
\begin{equation}
\frac{dp_i(t)}{dt}=\sum_{j} \left[ W_{ij}(t) p_j(t)-W_{ji}(t)p_i(t)\right]\, .
\label{2.1}
\end{equation}
Here $p_i(t)$ is the probability of finding the system in state $i$ at time $t$,
and the rates $W_{ij}$ for  transition from state $j$ to state $i$ depend on
time in a given way, independently of the state of the system. Let us define a
function
\begin{equation}
H(t)=\sum_i p_i(t) \ln \frac{p_i(t)}{p_i^\prime (t)}    \, ,
\label{2.2}
\end{equation}
where $p_i(t)$ and $p_i^\prime(t)$ are two solutions of Eq.\ (\ref{2.1})
corresponding to different initial conditions. The above definition for $H(t)$
assumes that $p_i^\prime(t)$ is positive for all the states $i$. This condition
will be fulfilled after a transient period if the process defined by Eq.\
(\ref{2.1}) is irreducible, even in the case that some initial probabilities
vanish. The time variation of $H(t)$ is given by
\begin{equation}\label{2.3}
\frac{dH(t)}{dt}=A(t)   \,
\end{equation}
with $A(t)$ being a complicated functional of the two solutions $p$ and
$p^\prime$ \cite{ByP93}, that can be written in the form
\begin{equation}\label{2.3a}
  A(t)=\sum_{ij} W_{ij} p_j^\prime \left[ \left( \frac{p_j}{p_j^\prime}-
  \frac{p_i}{p_i^\prime} \right) \left( \ln\frac{p_i}{p_i^\prime} +1 \right)
  +\frac{p_i}{p_i^\prime} \ln \frac{p_i}{p_i^\prime}
  -\frac{p_j}{p_j^\prime} \ln \frac{p_j}{p_j^\prime} \right]    \, .
\end{equation}
Its main property is that $A(t)\leq 0$. Besides, if the transition rates define
an irreducible process at time $t$, the equality sign holds only when
\begin{equation}\label{2.4}
p_i(t)=p_i^\prime (t)   \, ,
\end{equation}
for all the states $i$. As the function $H(t)$ is bounded below, $H(t)\geq 0$,
it must tend to a limit and, therefore,
\begin{equation}\label{2.5}
  \lim_{t\rightarrow\infty}p_i(t)=\lim_{t\rightarrow\infty}p_i^\prime(t)\, .
\end{equation}
Thus all the solutions  of the master equation converge toward the same
behavior, if the long-time limit of the transition rates still define an
irreducible process. This equation can be understood as showing the existence of
a long-time regime, where the influence of the initial conditions has been lost,
and the state of the system and its dynamics is fully determined by the law of
variation of the transition rates. Therefore, there will be a special solution
of the master equation such that all the other solutions approach it after an
initial transient period. We will refer to this special solution as the
``normal'' solution of the master equation for the given time dependence of the
transition rates. A more detailed and general discussion of the H-theorem
leading to the existence of this ``normal'' solution can be found in Ref.\
\cite{ByP93}.

Now suppose that the master equation models the dynamical behavior of a granular
system submitted to vertical vibration. The transition rates $W_{ij}$ would be
functions of the parameter $\Gamma$ characterizing the strength of the
vibration. If the granular pile is vibrated with sinusoidal pulses of amplitude
$A$ and frequency $\omega$, the quantity $\Gamma=A\omega^2/g$
\cite{KFLJyN95,NKBJyN98}, where $g$ is the gravity, is usually defined. In the
case of time dependent intensity $\Gamma$, the equation determining the
evolution of the granular pile will be of the type given by Eq.\ (\ref{2.1}).
Assuming that for arbitrary $\Gamma\neq 0$ the tapping process allows the system
to explore the whole configuration space, the stochastic process will be
irreducible and the existence of a normal solution for a given program of
variation of the intensity $\Gamma$ follows, provided that $\Gamma$ does not
vanish in the long-time limit.

Let us assume that for every value of the intensity $\Gamma>0$, the equation
\begin{equation}\label{2.6}
  \sum_j W_{ij}(\Gamma) p_{j}^{(s)}(\Gamma)=\sum_j W_{ji}(\Gamma)
  p_{i}^{(s)}(\Gamma)
\end{equation}
has a ``canonical'' solution, of the form
\begin{equation}\label{2.7}
  p_{i}^{(s)}(\Gamma)=\frac{1}{Z(X)} \exp \left[ -\frac{V_i}{\lambda X} \right]
    \, ,
\end{equation}
where $V_i$ is the volume of the system in state $i$,
\begin{equation}\label{2.8}
  Z(X)=\sum_i \exp \left[ -\frac{V_i}{\lambda X} \right]
\end{equation}
is a partition function, $\lambda$ is a constant with the dimension of volume,
and $X$ is a variable termed the compactivity of the granular system in the
framework of Edward's statistical mechanics theory of powders
\cite{EyO89,MyE89}. Of course, in the context considered here the compactivity
$X$ will be a function of the vibration intensity $\Gamma$, $X=f(\Gamma)$, which
has to be found  for each particular model. Equation (\ref{2.7}) defines the
steady state  reached by the system if the vibration intensity is constant in
time. The macroscopic value of the volume in this state is
\begin{equation}\label{2.9}
  \overline{V}^{(s)}(X)=\sum_i V_i \, p_{i}^{(s)}\left[ \Gamma(X) \right]  \, ,
\end{equation}
and the configurational entropy $S(X)$ reads
\begin{equation}\label{2.9b}
  S(X)=-\lambda \sum_i p_i^{(s)}(X) \ln p_i^{(s)}(X)    \, .
\end{equation}
Of course, the compactivity $X$ can be obtained from its usual definition,
\begin{equation}\label{2.9c}
  X=\frac{d\overline{V}^{(s)}}{dS}      \, .
\end{equation}
For the sake of simplicity, in the following we will take the unit of volume
such that $\lambda=1$. The stationary volume $\overline{V}^{(s)}(X)$ is an
increasing function of the compactivity $X$, since the powder
``compressibility''
\begin{equation}\label{2.10}
  \kappa (X)\equiv\frac{d \overline{V}^{(s)}(X)}{dX}=\frac{1}{X^ 2}
  \sum_i \left[ V_i-\overline{V}^{(s)}(X) \right]^2 p_{i}^{(s)}(X) \, ,
\end{equation}
is proportional to the volume fluctuations over the ensemble of granular systems
considered. Then, the compactivity $X$ should be an increasing function of the
vibration intensity $\Gamma$, because it is reasonable to expect that the powder
became fluffier with higher vibration intensities.  This is, for instance, the
behavior found in the simple two-volume model considered in Ref.\ \cite{EyG98}.
In general, this property must be checked once the relationship between $\Gamma$
and $X$, $X=f(\Gamma)$, has been derived  for each specific model.

Of course, $p_{i}^{(s)}$ is not a solution of the master equation when the
intensity $\Gamma$ is time-dependent, and in general the system does not
monotonically approach the steady distribution. Nevertheless, define [compare
with Eq.\ (\ref{2.2})]
\begin{equation}\label{2.12}
  H^{(s)}(t)=\sum_i p_i(t) \ln \frac{p_i(t)}{p_{i}^{(s)}(X)}      \, ,
\end{equation}
where $p_i(t)$ is again one solution of Eq.\ (\ref{2.1}), and $p_{i}^{(s)}(X)$
depend on time through the compactivity $X=X(t)$. If we define a statistical
entropy as \cite{ByP90}
\begin{equation}\label{2.12b}
  S^{*}(t)=-\lambda \sum_i p_i(t) \ln p_i(t)      \, ,
\end{equation}
and use the notation
\begin{equation}\label{2.12c}
  \overline{V}(t)=\sum_i V_i \, p_i(t)        \, ,
\end{equation}
for the actual average volume at time $t$, after a very simple algebra it is
found that
\begin{equation}\label{2.12d}
  H^{(s)}(t)=\frac{1}{X} \left[ \Bigl( \overline{V}(t)-X S^*(t) \Bigr) -
                \Bigl( \overline{V}^{(s)}(X)-X S(X) \Bigr) \right]     \, .
\end{equation}
Thus $H^{(s)}(t)$ is proportional to the deviation of the actual ``effective''
volume \cite{EyO89,MyE89} at time $t$,
\begin{equation}\label{2.12e}
  Y(t)=\overline{V}(t)-X S^{*}(t)  \, ,
\end{equation}
from its stationary value.

The time variation of $H^{(s)}$ is easily obtained as \cite{ByP93}
\begin{equation}\label{2.13}
  \frac{dH^{(s)}}{dt}=A^{(s)}(t)-\sum_i \frac{p_i(t)}{p_{i}^{(s)}(X)}
  \frac{dp_{i}^{(s)}(X)}{dX} \frac{dX}{dt} \, ,
\end{equation}
where $A^{(s)}(t)$ is given by Eq.\ (\ref{2.3a}), but replacing $p_i^\prime(t)$
by $p_{i}^{(s)}(X)$. Taking into account Eqs.\ (\ref{2.7}-\ref{2.9}), it is
found
\begin{equation}\label{2.14}
  \frac{dH^{(s)}}{dt}=A^{(s)}(t)-\frac{1}{X^2} \frac{dX}{dt}
  \left[ \overline{V}(t)-\overline{V}^{(s)}(X) \right]     \, .
\end{equation}
Equation (\ref{2.14}) does not have a well-defined sign and, therefore, the
stationary distribution is not monotonically approached, in general, when the
vibration intensity is time-dependent. However, in those processes such that the
compactivity (or, equivalently, the vibration intensity $\Gamma$) increases
monotonically in time, only the term
\begin{equation}\label{2.16}
  B(t)=\frac{1}{X^2}\frac{dX}{dt} \overline{V}^{(s)}(X)
\end{equation}
is positive. Using the analogy between temperature and compactivity, we will
refer to those processes as ``heating'' processes. If in  a given ``heating''
process it is verified that
\begin{equation}\label{2.17}
  \lim_{t\rightarrow\infty} B(t)=0      \, ,
\end{equation}
we can conclude
\begin{equation}\label{2.18}
  \lim_{t\rightarrow\infty} \frac{dH^{(s)}(t)}{dt} \leq 0  \, .
\end{equation}
Since $H^{(s)}(t)$ is bounded below, the only possibility is in fact the
equality, which implies
\begin{equation}\label{2.19}
  \lim_{t\rightarrow\infty} A^{(s)}(t)=0
\end{equation}
and
\begin{equation}\label{2.20}
  \lim_{t\rightarrow\infty} \frac{1}{X^2} \frac{dX}{dt} \overline{V}(t)
        =0 \, .
\end{equation}
From Eq.\ (\ref{2.19}) and the irreducibility of the stochastic process, it
follows that in the long time limit,
\begin{equation}\label{2.21}
  \lim_{t\rightarrow\infty} p_i(t)=\lim_{t\rightarrow\infty}p_{i}^{(s)}(X)
  \, ,
\end{equation}
i.\ e., the system goes to the steady curve for long enough times, if the
``heating'' program verifies the condition given by Eq.\ (\ref{2.17}).

The following picture emerges for the evolution of a granular system submitted
to vibrations with increasing intensity: Starting from an arbitrary initial
condition, in a first step the system tends to a behavior which is determined by
the law of variation of the vibration intensity $\Gamma$, and the initial
condition has been forgotten. This implies the existence of a special solution
of the master equation, called the ``normal'' solution, that is approached by
all the other solution after an initial transient period. Besides, if the system
is ``heated'' slowly, in the sense that Eq. (\ref{2.17}) is verified, the normal
solution tends afterwards to the stationary curve. This picture is similar to
the one found in some models of structural glasses \cite{ByP94,ByP93,PByS97}.

\section{Hilbert's method around the steady curve}
\label{seca}

In this Section we will use Hilbert's method to derive a quite general form of
the normal solution of Eq.\ (\ref{2.1}) near the steady state curve. We will
focus on the class of models for granular media considered in the previous
section, but the results can be directly extended to any system with a
``canonical'' distribution describing the stationary state.

Let us consider Eq. (\ref{2.1}), rewritten in the form \cite{vk92}
\begin{equation}\label{a1}
  \frac{d\bbox{p}(t)}{dt}=\bbox{\hat{W}}(t) \bbox{p}(t)        \, ,
\end{equation}
where $\bbox{p}$ is a vector (column matrix) whose elements are the
probabilities $p_i(t)$ of the $i$-th state of the system at time $t$, and
$\bbox{\hat{W}}$ is a square matrix with elements $\hat{W}_{ij}$ given by
\begin{equation}\label{a1b}
 \hat{W}_{ij}(t)=W_{ij}(t)-\delta_{ij} \sum_{k} W_{kj}(t)        \, .
\end{equation}
If the transition rates are time independent and the  detailed balance condition
is verified, solving Eq.\ (\ref{a1}) is equivalent to obtain the solution of the
eigenvalue problem
\begin{equation}\label{a1c}
  \bbox{\hat{W}} \bbox{\varphi}(q)=-\lambda (q) \bbox{\varphi}(q)  \, ,
\end{equation}
with $\lambda(q)>0$ for all $q$. The eigenvectors $\bbox{\varphi}(q)$ are
completed with the stationary distribution $\bbox{p}^{(s)}$, which is an
eigenvector of $\bbox{\hat{W}}$ corresponding to the null eigenvalue, i.\ e.,
\begin{equation}\label{a1d}
  \bbox{\hat{W}} \bbox{p}^{(s)}=0       \, .
\end{equation}
Besides,  if the Markov process is irreducible there is only one stationary
distribution with all its components positive \cite{vk92}. The matrix
$\bbox{\hat{W}}$ is hermitian with the following definition for the scalar
product of any two vectors $\bbox{a}$ and $\bbox{b}$,
\begin{equation}\label{a1e}
  (\bbox{a},\bbox{b})=\sum_i \frac{a_i b_i}{p_i^{(s)}}   \, .
\end{equation}

If the Markov process is irreducible and the detailed balance condition is
fulfilled for all times, Eqs.\ (\ref{a1c}-\ref{a1e}) remain valid for time
dependent transition rates. Of course, the eigenvalues and eigenvectors will
depend on time in general. The usual situation is that the transition rates
depend on time through an externally controlled parameter like the temperature
in a thermal system or the compactivity $X$ in a granular medium. Thus, we will
write sometimes in the following $\bbox{\hat{W}}(X)$, $\lambda (q,X)$,
$\bbox{\varphi} (q,X)$ and $\bbox{p}^{(s)}(X)$.

Hilbert's method consists  in solving the master equation by means of the
iterative process
\begin{mathletters}\label{a2}
\begin{equation}\label{a2a}
 \bbox{\hat{W}}(t) \bbox{p}^{(0)}(t)=0      \, ,
\end{equation}
\begin{equation}\label{a2b}
  \bbox{\hat{W}}(t) \bbox{p}^{(n)}(t)=\frac{d\bbox{p}^{(n-1)}(t)}{dt}
  \, ,    \;\;      n \geq 1    \, .
\end{equation}
\end{mathletters}
In this way we obtain a probability distribution
\begin{equation}\label{a3}
  \bbox{p}_H(t)=\sum_{n=0}^{\infty} \bbox{p}^{(n)}(t) \, ,
\end{equation}
which is a solution of Eq.\ (\ref{a1}). The first term $\bbox{p}^{(0)}(t)$ gives
the ``stationary'' distribution,
\begin{equation}\label{a4}
  p_i^{(0)}(t)=p_i^{(s)}(X)      \, ,
\end{equation}
where $X$ stands for the value of the compactivity at time $t$, $X=X(t)$. The
Hilbert solution $\bbox{p}_H$,  constructed following the above rules, is a
``normal'' solution, because it only depends on the external law of variation of
the transition rates, and it does not refer to any specific initial conditions.
Nevertheless, the range of validity of $\bbox{p}_H(t)$ is limited in general
because of  the divergence of Hilbert's expansion \cite{Resibois}. From a
physical point of view, this divergence is connected to the fact that we are
expanding $\bbox{p}(t)$, which may describe a situation arbitrarily far from the
steady state, around the stationary solution $\bbox{p}^{(s)}(X)$.

In ``heating'' processes we have shown in Sec.\ \ref{sec2} that a normal
solution of the master equation exists and, under very general conditions, it
tends to the stationary solution for very high compactivity. Thus in the high
compactivity limit Hilbert's method can be useful, since it is an expansion
around the steady state. We will restrict ourselves to the first correction
$\bbox{p}^{(1)}(t)$, because the difference between the probability
distributions $\bbox{p}_H(t)$ and $\bbox{p}^{(s)}(X)$ is expected to be small in
the high compactivity limit. The normalization of $\bbox{p}^{(s)}(X)$ implies
that
\begin{equation}\label{a4b}
  \left( \bbox{p}^{(s)}(X),\frac{d\bbox{p}^{(s)}(X)}{dX} \right)=\sum_i
  \frac{dp_i^{(s)}(X)}{dX}=0       \, ,
\end{equation}
and, therefore,
\begin{equation}\label{a5}
  \bbox{p}^{(1)}(t)=\bbox{\hat{\cal T}}(X) \frac{d\bbox{p}^{(s)}(X)}{dX}
  \frac{dX}{dt}\, ,
\end{equation}
where $\bbox{\hat{\cal T}}(X)$ is the inverse operator of $\bbox{\hat{W}}(X)$ in
the space orthogonal to the equilibrium distribution $\bbox{p^{(s)}}(X)$. Using
Eq.\ (\ref{2.7}), we obtain
\begin{equation}\label{a7}
  \frac{dp_i^{(s)}(X)}{dX}=\frac{1}{X^2} \left[ V_i-
  \overline{V}^{(s)}(X) \right] p_i^{(s)}       \, .
\end{equation}
By introducing a function
\begin{equation}\label{a8}
  \xi (q,X)=\sum_{i} V_i \varphi_i (q,X)      \, ,
\end{equation}
where $\bbox{\varphi}(q,X)$ is the eigenvector defined in  Eq.\ (\ref{a1c}), we
can rewrite Eq.\ (\ref{a7}) in a vectorial form as
\begin{equation}\label{a9}
  \frac{d\bbox{p}^{(s)}(X)}{dX}=\frac{1}{X^2} \sum_q \xi(q,X)
  \bbox{\varphi}(q,X) \, .
\end{equation}
The above expression follows from the completeness of the eigenvectors and the
property
\begin{equation}\label{a9b}
  \left( \bbox{\varphi}(q,X),\frac{d\bbox{p}^{(s)}(X)}{dX} \right)=
  \frac{1}{X^2} \sum_i \varphi_i (q,X)
  \left[ V_i-\overline{V}^{(s)}(X) \right]=\frac{1}{X^2} \xi(q,X)  \, .
\end{equation}
The term proportional to $\overline{V}^{(s)}(X)$ vanishes as a consequence of
the orthogonality of the eigenvectors $\bbox{p}^{(s)}(X)$ and
$\bbox{\varphi}(q,X)$ for all $q$. Then, to first order in deviations from the
equilibrium curve we have
\begin{equation}\label{a10}
  \bbox{p}_H(t)=\bbox{p}^{(s)}(X)-\frac{1}{X^2} \frac{dX}{dt} \sum_q
  \lambda^{-1}(q,X) \xi(q,X) \bbox{\varphi}(q,X) \, .
\end{equation}
The notation used reflects the fact that the right hand side  depends on time
only through the compactivity $X$ of the system.

From Eq.\ (\ref{a10}) it is possible to evaluate the average values of the
physical quantities of the system over Hilbert's distribution in the first order
approximation. For instance, the mean value of the volume is
\begin{equation}\label{a11}
  \overline{V}_H (t)=\overline{V}^{(s)}(X)-\frac{1}{X^2} \frac{dX}{dt} \sum_q
  \lambda^{-1}(q,X) \xi^2(q,X)      \, ,
\end{equation}
This expression can be written in a more transparent way by using the following
identity for the powder ``compressibility'' defined in Eq.\ (\ref{2.10}),
\begin{equation}\label{a13}
  \kappa(X)=\frac{1}{X^2} \sum_q \xi^2(q,X)        \, ,
\end{equation}
and the expression for the average relaxation time of the volume found in linear
relaxation theory (see the appendix)
\begin{equation}\label{a14}
  \tau(X)=\frac{\sum_q \lambda^{-1}(q,X) \xi^2(q,X)}{\sum_q \xi^2(q,X)} \, .
\end{equation}
In this way, it is found
\begin{equation}\label{a15}
  \overline{V}_H(t)=\overline{V}^{(s)}(X)-\frac{dX}{dt}\kappa(X)\tau(X) \, .
\end{equation}
It is important to note that the detailed balance condition is a key point in
our derivation of Eq.\ (\ref{a15}), while the existence of the normal solution
does not need detailed balance to be satisfied, but it follows if the Markov
process is irreducible in the long time limit. On the other hand, Eq.\
(\ref{a15}) can also be applied to estimate the departure from the stationary
curve in ``cooling'' processes, if the system is initially prepared in the
stationary state. In other words, Eq.\ (\ref{a15}) should be applicable to any
situation near the stationary curve. Therefore, it allows to calculate the first
order in the deviation of the volume from its steady value over the normal curve
when the system asymptotically tends to the stationary curve, as it is the case
of ``heating'' processes. The same remark can be made about Eq.\ (\ref{a10}) for
Hilbert's distribution $\bbox{p}_H(t)$. It contains all the information needed
to evaluate one-time properties over the normal curve in situations close to the
steady state.

\section{The model}
\label{sec3}

In this Section we will briefly  review a simple model for the
vibrocompaction of a dry granular system which has been recently introduced
\cite{BPyS99,BPyS98a}. We consider a one-dimensional lattice with $N$ sites.
Each site can be either occupied by a particle or empty, i.\ e., occupied by a
hole, with the restriction that there cannot be two nearest neighbor holes (such
a configuration would be unstable). A variable $m_i$ is assigned to each site
$i$, taking the value $m_i=1$ if the site is empty, and  $m_i=0$ if there is a
particle on it. Then, a configuration of the system is fully specified by giving
the values of the set of variables $\bbox{m}\equiv
\{m_1,m_2,\cdots,m_N\}.$

The dynamics of the system is defined as a Markov process, and formulated by
means of a master equation for the probability $p(\bbox{m},t)$ of finding the
system in the configuration $\bbox{m}$ at time $t$ \cite{BPyS99},
\begin{equation}\label{3.1}
  \frac{d}{dt} p(\bbox{m},t)= \sum_{\bbox{m}'}
   \left[ W(\bbox{m}|\bbox{m}')
  p(\bbox{m}',t)-W (\bbox{m}'|\bbox{m}) p(\bbox{m},t) \right]      \, ,
\end{equation}
where $W(\bbox{m}|\bbox{m}')$ is the transition rate from state $\bbox{m}'$ to
state $\bbox{m}$. The possible transitions in the system can be classified in
three groups. Firstly, there are transitions conserving the number of particles,
corresponding to purely diffusive processes. Their rates are given by
\begin{equation}\label{3.2}
        W(010|100)=W(010|001)=\frac{1}{2}\alpha \, ,
\end{equation}
with $\alpha$ being a constant. These transition rates must be understood as it
is usually done, as the transition rates between states connected by the given
rearrangement. Only the variables of the set of sites involved in the transition
are indicated in the notation. Secondly, there are also transitions increasing
the number of particles, with rates
\begin{mathletters}\label{3.3}
\begin{equation}\label{3.3a}
  W(010|101)=\frac{1}{2} \alpha  \, ,
\end{equation}
\begin{equation}\label{3.3b}
  W(001|101)=W(100|101)=\frac{1}{4} \alpha        \, .
\end{equation}
\end{mathletters}
Finally, the transition rates for those processes decreasing the number of
particles are
\begin{mathletters}\label{3.4}
\begin{equation}\label{3.4a}
  W(01010|00100)=\frac{1}{2} \alpha^2 \, ,
\end{equation}
\begin{equation}\label{3.4b}
  W(01010|01000)=W(01010|00010)=\frac{1}{4}\alpha^2      \, .
\end{equation}
\end{mathletters}
The transition rates in Eqs.\ (\ref{3.2})-(\ref{3.4}) define an effective
dynamics for tapping processes  taking place in a previous more general model
presented in Ref.\ \cite{BPyS98a}. The parameter $\alpha$ characterizes the
tapping process completely. Note that we have rescaled all the expressions of
the transition rates in Ref. \cite{BPyS99} in such a way that the time scale $t$
is a measure of the number of taps. For $\alpha=0$ no transition is possible in
the system and, therefore, the parameter $\alpha$ measures the intensity of the
vibration. The configuration with the highest density, i.\ e., no holes present,
is completely isolated from the rest of the states, in the sense that no
transition is possible from or towards it for any value of $\alpha$. Aside from
this particular state,  all the other possible configurations are connected
through a chain of transitions with non-zero probability for $\alpha\neq 0$, and
the Markov process defining the dynamics is then irreducible.

When $\alpha$ does not depend on time, i.\ e., the intensity of vibration is
constant, there is a unique stationary solution $p^{(s)}(\bbox{m})$ of the
master equation (\ref{3.1}). We will use the notation $\bbox{m^{(k)}}$ for a
configuration of the system with $k$ holes. Clearly, it is verified that $1\leq
k\leq (N+1)/2$. It is easily found that \cite{BPyS99}
\begin{equation}\label{3.5}
  p^{(s)}(\bbox{m^{(k)}})=\frac{e^{-k/X}}{Z}    \, ,
\end{equation}
where $Z$ is the normalization constant, and $X$ is related to the vibration
intensity $\alpha$ through the relation
\begin{equation}\label{3.6}
  \alpha=e^{-1/X}       \, .
\end{equation}
Comparison with Eq.\ (\ref{2.7}) shows that in our model the number of holes $k$
plays the role of the volume of the system and $X$ is the compactivity of
Edward's theory of powders \cite{EyO89,MyE89}. More precisely, $k$ would be
proportional to the excess volume from the densest state. Interestingly, the
compactivity $X$ is an increasing function of $\alpha$, as it has been argued on
general basis in Section \ref{sec2}. Besides, the compactivity vanishes for the
no vibrated case. The partition function can be analytically derived in the
infinite system limit, with the result \cite{BPyS99}
\begin{equation}\label{3.7}
  \ln\zeta \equiv \frac{1}{N} \ln Z=
  -\ln 2 +\ln \left[1+\left(1+4\alpha\right)^{1/2}\right] \, .
\end{equation}
From here, the stationary value of the density of holes is obtained in the
standard way,
\begin{equation}\label{3.8}
  x_1^{(s)}=\frac{\overline{k}^{(s)}}{N}=-\frac{d\ln\zeta}{d(1/X)}=
  \frac{1}{2} \left[ 1- \left( 1+4\alpha \right)^{-1/2} \right] \, .
\end{equation}
The stationary density of holes increases monotonically from the densest state
$x_1^{(s)}=0$ to the fluffiest state $x_1^{(s)}=1/2$ as the vibration intensity
increases monotonically from $\alpha=0$ to $\alpha=\infty$, since the analogous
to the  ``compressibility''
\begin{equation}\label{3.9}
  \kappa=\frac{dx_1^{(s)}}{dX}=
  \frac{e^{-1/X}}{X^2 \left( 1+4 e^{-1/X} \right)^{3/2}}
\end{equation}
is positive definite for any value of $X$. Nevertheless, the densest
configuration cannot be actually reached, since for $\alpha=0$ no transition is
possible and the system gets trapped in its initial state. As the number of
holes $k$ is an upper bounded variable, the compactivity $X$ can take negative
values, corresponding to configurations fluffier than those of positive
compactivities. For instance, $X\rightarrow 0^{-}$ gives
$\alpha\rightarrow\infty$ and $x_1^{(s)}=1/2$, the least dense state.

Finally, let us qualitatively study the dynamics of the system for
time-independent transition rates in the low vibration limit, $\alpha\ll 1$ or
$X\rightarrow 0^{+}$. In that limit, the stationary value of the density of
holes $x_1^{(s)}$ is very small, and a typical configuration of the system
consist of a few holes, separated by long arrays of particles. Since
$x_1^{(s)}\sim\alpha$, the mean distance between holes is of order
$\alpha^{-1}$. Moreover, at least in linear relaxation theory and at low
compactivities, the evolution  of the system will be mainly associated to
diffusive processes. The characteristic time $\widehat{\tau}$ of the relaxation
process would be the square of the mean distance between holes divided by an
effective diffusion coefficient, which can be estimated from Eq.\ (\ref{3.2}) as
of the order $\alpha$. Therefore,
\begin{equation}\label{3.10}
  \widehat{\tau} \propto \alpha^{-3}      \, .
\end{equation}
This result will be important in the following, since an estimate of the
relaxation time in the linear relaxation regime is necessary when studying the
time-dependent rates case, as it follows from Eq.\ (\ref{a15}).

\section{Processes with time-dependent vibration intensity}
\label{sec4}

Next we will study the dynamical behavior of the model  described in the
previous section, when submitted to processes in which the vibration intensity
$\alpha$ changes in  time in a given way. Due to the relationship between
$\alpha$ and the compactivity $X$, Eq.\  (\ref{3.6}), this is equivalent to
consider that the compactivity $X$ varies in  time following a given law. As
already mentioned, we will refer to a process as a ``heating'' (``cooling'') one
when the vibration intensity is monotonically increased (decreased). Such kind
of processes have been already considered in the literature, both in real
granular systems \cite{NKBJyN98,Ja98} and in simple models \cite{NyC99}.

The general results of Sec.\ \ref{sec2} are applicable to this particular model.
The only limitation is due to  the loss of irreducibility of the dynamics for
$\alpha=0$. Thus the existence of a special solution of the master equation,
such that all the others approach it in the long time limit, applies to any
program of variation of $\alpha$  except for ``cooling'' processes  up to $X=0$.
As a consequence, for ``heating'' processes there will be a special ``normal''
curve, to which all the other solutions tend at a first stage. Later on, the
system will approach the ``stationary'' distribution $p^{(s)}[\alpha(t)]$ in the
long time limit, provided that the condition in Eq.\ (\ref{2.17}) is verified,
i.\ e.,
\begin{equation}\label{4.1}
  \lim_{t\rightarrow\infty} \frac{1}{X^2(t)} \frac{dX(t)}{dt}
  x_1^{(s)}\left[ \alpha (t) \right]=0    \, .
\end{equation}
Taking into account that $x_1^{(s)}$ is upper bounded by $1/2$, and the
relationship between $\alpha$ and $X$, Eq.\ (\ref{3.6}), the above condition can
also be written as
\begin{equation}\label{4.3}
  \lim_{t\rightarrow\infty} \frac{d\ln\alpha(t)}{dt}=0  \, .
\end{equation}
Equation (\ref{4.3}) expresses a restriction for the ``heating'' programs
driving the system to the stationary curve in the long time limit, but it does
not affect at all the existence of the normal solution, which only depends on
the ergodicity of the process.

Application of Eq.\ (\ref{a15})  to the present model yields
\begin{equation}\label{4.5}
  x_1(t)=x_1^{(s)}[X(t)]-\frac{dX(t)}{dt} \kappa[X(t)] \tau[X(t)]+\cdots \, ,
\end{equation}
where $\tau(X)$ is the mean relaxation time of the density in the linear
relaxation approximation. Upon writing the above expression we have taken into
account that in our model the number of holes plays the role of the volume. For
high compactivities the second term on the right hand side  of Eq.\ (\ref{4.5})
is negligible against the first one, since the compressibility
$\kappa\rightarrow 0$ when $X\rightarrow\infty$. Then, for high vibration
intensities the system remains over the stationary curve. As the compactivity
decreases, the system departs from that curve, as a consequence of the increase
of the mean relaxation time $\tau$, which is expected to be proportional to the
characteristic time $\widehat{\tau}$ estimated in Eq.\ (\ref{3.10}).

There is some freedom when choosing the law of variation of the vibration
intensity $\alpha$. Our choice will be  motivated by simplicity, but also by the
analogies with a glass-like behavior previously found in granular systems
\cite{Ja98,NyC99}. In supercooled liquids the temperature is usually varied at a
constant rate \cite{Sch86}, so we have considered processes in which the
compactivity changes linearly in time,
\begin{equation}\label{4.4}
  \frac{dX}{dt}=\pm r   \, ,
\end{equation}
with $r>0$, which is equivalent to
\begin{equation}\label{4.4b}
  \frac{d\alpha}{dt}=\pm r \alpha \left( \ln\alpha \right)^2 \, ,
\end{equation}
the plus sign corresponding to ``heating'' processes and the minus sign to
``cooling'' programs.

 The rest of this section is organized as follows. First, we study
``cooling'' processes. The system is initially put in  the stationary state
corresponding to a given value $\alpha_0$ of the vibration intensity. Then, the
compactivity $X=-1/\ln\alpha$ is decreased following Eq. (\ref{4.4}). The
existence of a phenomenon analogous to the laboratory glass transition of
supercooled liquids arises. Afterwards, ``heating'' processes are considered,
paying special attention to the appearance of hysteresis effects, and relating
them to the trend of the system to approach the normal curve.

\subsection{``Cooling'' processes}
\label{sec4a}

We consider the continuous decreasing of the compactivity of the system from a
given initial value $X_0$ down to $X=0$. The latter corresponds to $\alpha=0$,
i.\ e., no vibration. The system is initially placed  in the stationary state
corresponding to the value $\alpha_0=\exp(-1/X_0)$ of the vibration intensity.
Then, the compactivity is decreased following the law
\begin{equation}\label{4.6}
  \frac{dX}{dt}=-r_c    \, .
\end{equation}

Our starting point will be Eq.\ (\ref{4.5}), particularized for the ``cooling''
process we are considering, i.\ e.,
\begin{equation}\label{4.7}
  x_1(t)=x_1^{(s)}[X(t)]+r_c \, \kappa[X(t)] \, \tau[X(t)]+O(r_c^2)    \, .
\end{equation}
As we have already discussed, from the above equation follows that the system
remains in equilibrium at ``high'' compactivities, since the second term in the
right hand side vanishes for $X\rightarrow\infty$. Nevertheless, as the
compactivity becomes smaller this term grows, due to the increase of the
relaxation time $\tau$. This means that there will exist a range of values of
the compactivity in which the second term is comparable to the first one. A
rough estimate of the value of the compactivity $X_g$ at which the system would
depart from the stationary curve can be obtained by equalling both terms. If we
consider that the system is slowly ``cooled'', $r_c\ll 1$, it is also
$\alpha_g=e^{-1/X_g} \ll 1$, and we can approximate both terms for their leading
behaviors in that limit, namely
\begin{mathletters}\label{4.8}
\begin{equation}\label{4.8a}
  x_1^{(s)}(X) \sim \alpha  \, ,
\end{equation}
\begin{equation}\label{4.8b}
  \kappa(X)= \frac{dx_1^{(s)}}{dX} \sim \alpha \left( \ln\alpha \right)^2   \,
\end{equation}
\begin{equation}\label{4.8c}
  \tau(X) \sim \tau_0 \alpha^{-3}        \,
\end{equation}
\end{mathletters}
where $\tau_0$ is a constant of the order of unity. Therefore, we get
\begin{equation}\label{4.10}
  r_c \tau_0 \left(\ln\alpha_g\right)^2= \alpha_g^3     \, .
\end{equation}
For $r_c \ll 1$ it is
\begin{equation}\label{4.11}
  \alpha_g \sim r_c^{1/3} \left| \ln r_c \right|^{2/3}       \, .
\end{equation}
We have omitted the factors containing $\tau_0$, as well as any other factor of
the order of unity.

For $\alpha < \alpha_g$ the system is effectively ``frozen'', due to the
divergent tendency of the relaxation time no more transitions are possible, and
the density of holes will be  approximately constant in this region. A measure
of the effective number of transitions left to the system before reaching
$\alpha=0$ from a given time $t$ is given by the scale \cite{ByP94,PByS97}
\begin{equation}\label{4.13}
  s(t)=\int_{t}^{t_0} dt' \frac{1}{\tau[X(t')]}    \, ,
\end{equation}
where $t_0$ is the time instant for which the ``cooling'' program finishes,
i.e.\,  $\alpha(t_0)=0$. In that way, the system would get frozen for $t>t_f$
such that $s(t_f)=1$, being easily obtained that $\alpha(t_f)\simeq \alpha_g$.
Using  Eq.\ (\ref{4.11}) it is possible to estimate the leading order value of
the compactivity at which the system gets frozen,
\begin{equation}\label{4.14}
  X_g=-\frac{1}{\ln\alpha_g} \sim \frac{3}{|\ln r_c |}     \, .
\end{equation}
This kind of behavior has been also found numerically in a granular system model
\cite{NyC99}. The inverse logarithmic dependence of the cooling rate is typical
for the laboratory glass transition temperature of supercooled liquids
\cite{Sch86}, and it has also been analytically derived in some simple models of
structural glasses \cite{ByP94,PByS97}.

Since the density of holes remains nearly constant for $\alpha<\alpha_g$, and
the two first terms of Hilbert's expansion (\ref{4.7}) are of the same order for
$\alpha\simeq\alpha_g$, it is reasonable to expect that
\begin{equation}\label{4.12}
  x_{1,\mbox{res}}=\lim_{t\rightarrow t_0} x_1(t) \propto x_1^{(s)}({\alpha_g})
  = \alpha_g \sim r_c^{1/3} \left| \ln r_c \right|^{2/3}      \, ,
\end{equation}
where $x_{1,\mbox{res}}$ is the residual value of the density of holes,
extending again to this model of granular system the terminology of  structural
glasses.

In order to check the above results, Fig.\ \ref{fig1} shows the residual value
of the density of holes as a function of the ``cooling'' rate, measured by the
parameter
\begin{equation}\label{4.15}
  \delta=r_c \left( \ln r_c \right)^2   \, ,
\end{equation}
in a log-log scale. The numerical  result agrees with the theoretical
prediction, Eq.\ (\ref{4.12}), since the curve is well fitted by a straight line
with a slope approximately equal to $1/3$. In Fig.\ \ref{fig1b} the evolution of
the density of particles, $\rho=1-x_1$, in a ``cooling'' process with rate
$r_c=10^{-5}$ is plotted. For comparison, the equilibrium curve, given by Eq.\
(\ref{3.8}), is also shown. The estimate of the freezing compactivity from Eq.\
(\ref{4.14}) is $X_g\simeq 0.26$, which is seen to be in good agreement with the
region in which the Monte Carlo density is approximately constant. Similar
behaviors have been observed for other small values of the cooling rate.

\subsection{``Heating'' processes and hysteresis effects}
\label{sec4b}

Let us analyze ``heating'' processes, i.\ e., processes in which the vibration
intensity is monotonically increased. In this  kind of processes, the dynamics
of the system is irreducible. Thus there is a ``normal'' solution of the master
equation, such that any other solution tends to it. Besides, if the heating
program verifies Eq.\ (\ref{4.3}), the normal solution approaches the stationary
curve for large enough times. We are going to discuss how these results can be
applied to our model, in order to understand its behavior when ``heated'' from
$\alpha=0$.

The compactivity will be increased according to the law
\begin{equation}\label{4.16}
  \frac{dX}{dt}=r_h     \, ,
\end{equation}
where $r_h$ is the rate for this process. According to Eq.\ (\ref{4.5}), the
system will approach the stationary curve in the high compactivity (long time)
limit. In the vicinity of the stationary curve, the evolution of the system is
given by
\begin{equation}\label{4.17}
  x_1(t)=x_1^{(s)}[X(t)]-r_h \kappa[X(t)] \tau[X(t)]+O(r_h^2)       \, .
\end{equation}
This equation explains why, in ``heating  processes'', the system tends to the
stationary curve following a curve different from the ``cooling'' one, even when
$r_c=r_h$. It follows directly from the comparison of Eqs.\ (\ref{4.17}) and Eq.
(\ref{4.7}), by noting that the deviation from the stationary behavior is of
opposite signs in ``heating'' and ``cooling'' processes. Therefore, hysteresis
effects show up.

However, perhaps the main result for ``heating'' processes is the existence of
the normal solution. Fig.\ \ref{fig2} shows the evolution of the density of
particles in a heating process with $r_h=10^{-5}$. The shape of the curve
depends on the initial condition for $X=0$. Two different initial preparations
of the system have been considered. The system was previously cooled down to
$X=0$, following two different linear programs with $r_c=10^{-5}$ and
$r_c=10^{-3}$, respectively. For the sake of clarity, these ``cooling'' curves
are not shown. From Fig.\  \ref{fig2} it is seen that both ``heating'' curves
tend to a common behavior and, afterwards, they approach the stationary curve
for high compactivities. Also plotted is the normal curve, which was obtained by
starting from the loosest packing state, $x_1=0.5$ \cite{stgl}.

Figure\ \ref{fig3} depicts a particular cycle of ``cooling'' and ``heating''
with the same rate, namely $r_c=r_h=10^{-5}$, as well as  the normal curve of
the ``heating'' process.  A behavior similar to the one found in real granular
systems \cite{NKBJyN98,Ja98}, and also in the ``Tetris'' model \cite{NyC99}, is
observed. When starting the heating process from the loosest packing state the
normal curve is obtained, which tends to the stationary behavior in the high
vibration intensity limit. Afterwards, cooling and reheating with the same rate
leads to the other two curves of the figure. These are approximately
``reversible'' for very small rates, since the deviation from the stationary
curve is smaller the smaller the rate. Nevertheless, they cannot be used to
obtain the stationary values of the density for low compactivities, due to the
glass-like kinetic transition. On the other hand, at high compactivities the
deviations of the system from the stationary curve for the ``cooling'' and the
``heating`` processes  are symmetric, as predicted by Eqs.\ (\ref{4.7}) and
(\ref{4.17}).

\section{Final remarks}
\label{sec5}

In the first part of the paper we have considered a wide class of models of
granular systems, namely those models whose dynamics under tapping  is described
by means of a master equation. It has been assumed that the tapping process is
such that the system is able to explore the whole space of metastable states,
when it is tapped with any vibration intensity $\Gamma\neq 0$. Then, if $\Gamma$
changes in time, and the Markov process remains irreducible in the long time
limit, all the solutions of the master equation tend to approach a special
solution, called the ``normal'' solution  for the given vibration program.
Besides, if the stationary distribution for constant $\Gamma$ is consistent with
Edward's statistical mechanics theory of powders, the normal solution tends to
the stationary curve in the high vibration intensity regime, provided that the
``heating'' program is not too fast, in the sense given by Eq.\ (\ref{2.17}). It
has also been argued that the compactivity $X$ of Edward's theory should be an
increasing function of the vibration intensity $\Gamma$. A quite general
prediction for the behavior of the normal solution in situations near the steady
state has been found, by using Hilbert's method to solve the master equation.
This result can be applied to any system having a ``canonical'' steady state
distribution.

Very recently, a simple model  for a granular system submitted to vertical
vibration with an stationary state described by Edward's theory of powders has
been introduced \cite{BPyS99,BPyS98a}. The model belongs to the class of systems
described in the previous paragraph, and here we have investigated its behavior
in processes in  which the vibration intensity depends on time. Such kind of
processes have been carried out in real granular systems \cite{NKBJyN98,Ja98},
and also investigated in some models \cite{NyC99}. For the sake of concreteness
we have taken the compactivity as a linear function of time.

The behavior of the model under ``cooling'' processes up to zero vibration
intensity exhibits a phenomenon similar to the laboratory glass transition. The
``cooling'' evolution departs from the stationary curve, and freezes, in a
narrow region around  the value of the compactivity $X_g$ such that the
effective number of transitions left to the system until reaching $X=0$ becomes
smaller than unity. This allows us to estimate the dependence of $X_g$ and the
residual values of the density upon the ``cooling'' rate. The results are
analogous to those obtained previously in simple models of structural glasses
\cite{ByP94,PByS97}.

In ``heating'' processes, a crucial role is played by the ``normal'' solution of
the master equation, which is completely determined by the program of variation
of the compactivity, and attracts any other solution, independently of the
initial condition. The hysteresis effects found when the system is ``cooled''
and ``reheated'' are related to the trend of the system to approach the normal
curve. The  normal curve corresponds to ``heating'' the system from the loosest
state. In a first stage, the system tends to approach monotonically the
``normal'' curve and, for longer times, corresponding to high enough
compactivities, the stationary curve is reached if the ``heating'' program
verifies Eq.\ (\ref{4.3}).

The work reported here suggests a very close relationship between structural
glasses and the behavior of a granular system under vibrations, supporting
previous results in the same direction \cite{BPyS98a,NyC99,Kad99}. It also
raises some questions, both from a theoretical and experimental point of view.
We think that it is worth checking the existence of the normal solution in real
granular systems, and whether it can be constructed starting from the loosest
packing state. Also, it seems interesting to know if the glass analogy also
extends to the linear relaxation regime, giving a stretched exponential behavior
of the response functions, or if the inverse logarithmic relaxation law
\cite{NKBJyN98} remains valid in such a region. The latter possibility would
lead to the need of investigating the reason for such a behavior.

\acknowledgments
This research was partially supported by the Direcci\'{o}n General de
Investigaci\'{o}n Cient\'{\i}fica y T\'{e}cnica (Spain) through Grant No.
PB98--1124.

\appendix

\section{Average relaxation time of the volume}

Linear relaxation describes the evolution of the system at constant value of the
compactivity $X$, when starting from the steady state corresponding to a
compactivity $X+\Delta X$, with $\Delta X$ very small, in the sense that we can
approximate
\begin{equation}\label{b1}
\bbox{p}^{(s)}(X+\Delta X)=\bbox{p}^{(s)}(X)+
\frac{d\bbox{p}^{(s)}(X)}{dX} \Delta X  \, .
\end{equation}
Taking into account Eq.\ (\ref{a9}), the evolution of the probability
distribution from this initial condition, $\bbox{p}(t=0)=\bbox{p}^{(s)}(X+\Delta
X)$, is
\begin{equation}\label{b2}
  \bbox{p}(t)=\bbox{p}^{(s)}(X)+\frac{\Delta X}{X^2} \sum_q \xi(q,X)
  \bbox{\varphi} (q,X) e^{-t\lambda(q,X)}       \, ,
\end{equation}
because $\bbox{\varphi}(q,X)$ is the eigenvector of the evolution operator
corresponding to the eigenvalue $\lambda(q,X)$. From the  above expression it is
straightforward to calculate the time evolution of the average volume,
\begin{equation}\label{b3}
  \overline{V}(t)-\overline{V}^{(s)}(X)=\frac{\Delta X}{X^2} \sum_q \xi^2(q,X)
                e^{-t\lambda(q,X)}      \, ,
\end{equation}
where we have made use of the definition of the function $\xi(q,X)$, Eq.\
(\ref{a8}).

The volume linear relaxation function corresponding to a value of the
compactivity $X$ is defined in the usual way,
\begin{equation}\label{b4}
  \phi_V(t;X)\equiv
  \frac{\overline{V}(t)-\overline{V}^{(s)}(X)}
       {\overline{V}(t=0)-\overline{V}^{(s)}(X)}        \, ,
\end{equation}
yielding
\begin{equation}\label{b5}
  \phi_V(t;X)=\frac{\sum_q \xi^2(q,X) e^{-t\lambda(q,X)}}
                  {\sum_q \xi^2(q,X)}   \, .
\end{equation}
On the other hand, the linear relaxation time  for the volume $\tau(X)$ is given
by definition by the area under the curve $\phi_V(t;X)$ as a function of $t$.
This leads to Eq.\ (\ref{a14}).

\begin{figure}
\caption{
Residual value $x_{1,\mbox{res}}$ of the hole density as a function of the
``cooling'' parameter $\delta$, obtained from Monte Carlo simulation of the
system. The continuous line is the best linear fit, having a slope equal to
$0.32$, very close to the theoretical value.}
\label{fig1}
\end{figure}

\begin{figure}
\caption{
Evolution of the particle  density $\rho$ with the compactivity $X$ for a
cooling process with rate $r_c=10^{-5}$ (diamonds). The  dotted line is the
equilibrium function. The estimate for the freezing compactivity $X_g$ is also
shown. }
\label{fig1b}
\end{figure}

\begin{figure}
\caption{
Evolution of the density of particles $\rho$ with the compactivity $X$ for a
heating process  with rate $r_h=10^{-5}$. Starting from the top, the symbols
correspond to heating processes after coolings with rates $r_c=10^{-3}$
(crosses) and $r_c=10^{-5}$ (diamonds). Both curves tend to approach the normal
curve (continuous line), whose initial condition is the loosest configuration,
$\rho=0.5$.}
\label{fig2}
\end{figure}

\begin{figure}
\caption{
Reversible-irreversible cycle, corresponding to a rate $r_c=r_h=10^{-5}$.
Starting from the loosest configuration, the system is heated (squares), tending
to the equilibrium curve (dotted line) for high enough compactivities.
Afterwards the system is cooled (diamonds) and reheated (crosses). The last two
processes are approximately reversible.}
\label{fig3}
\end{figure}


\begin{references}
\bibitem{stgl}
It can be easily understood why the usual choice for the initial condition made
in models of structural glasses \cite{ByP94,ByP93,PByS97} does not work in the
present context. In those models the initial condition was taken to be the limit
of the stationary curve for zero temperature, which corresponds here to the
densest state, $x_1=0$. But the densest configuration does not evolve for any
value of the compactivity, as it should also happen in a perfectly ordered real
granular system. Then, in our model the normal solution corresponds to putting
the system initially in the other extreme limit of the density, the loosest
configuration.
\end{references}
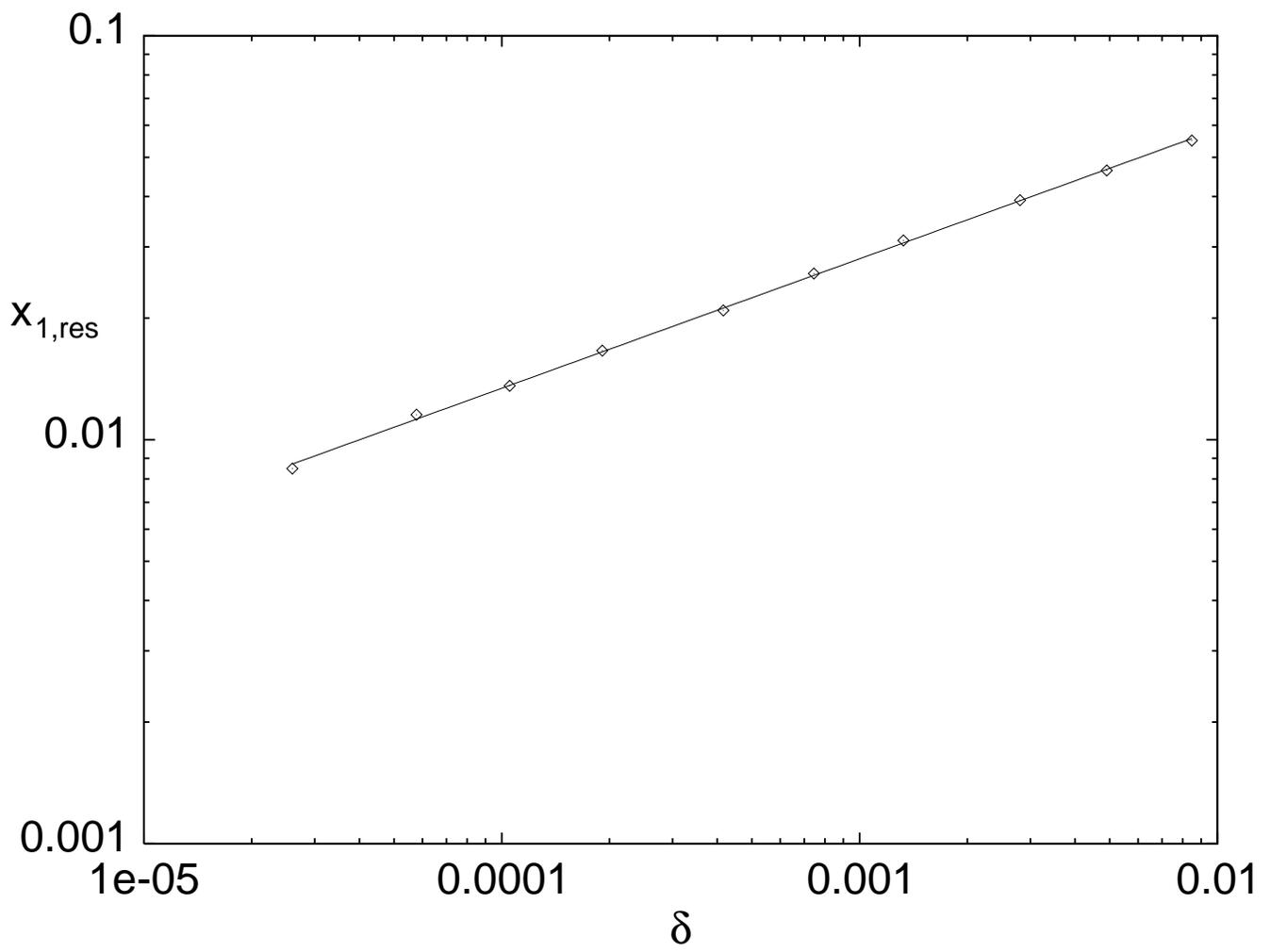
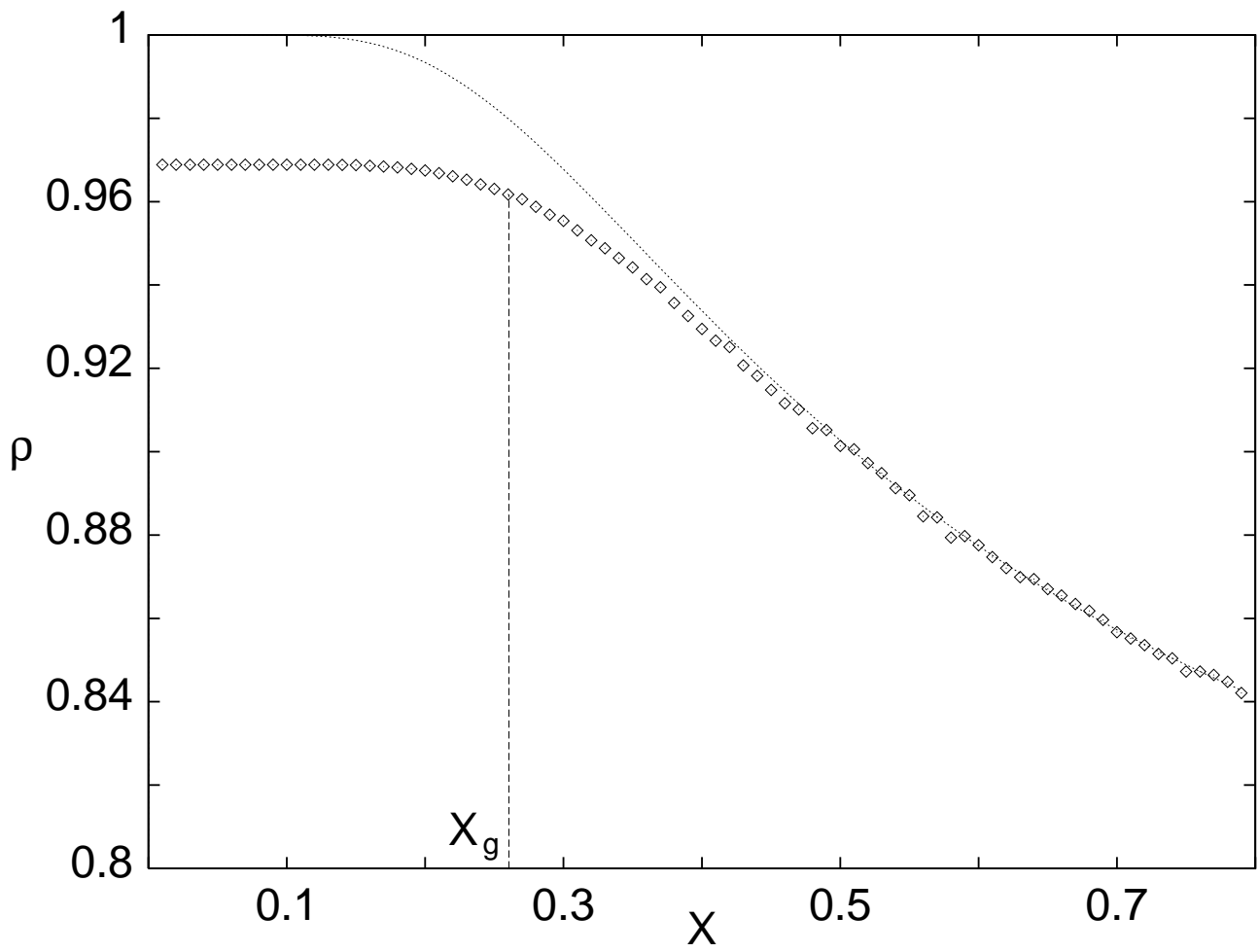
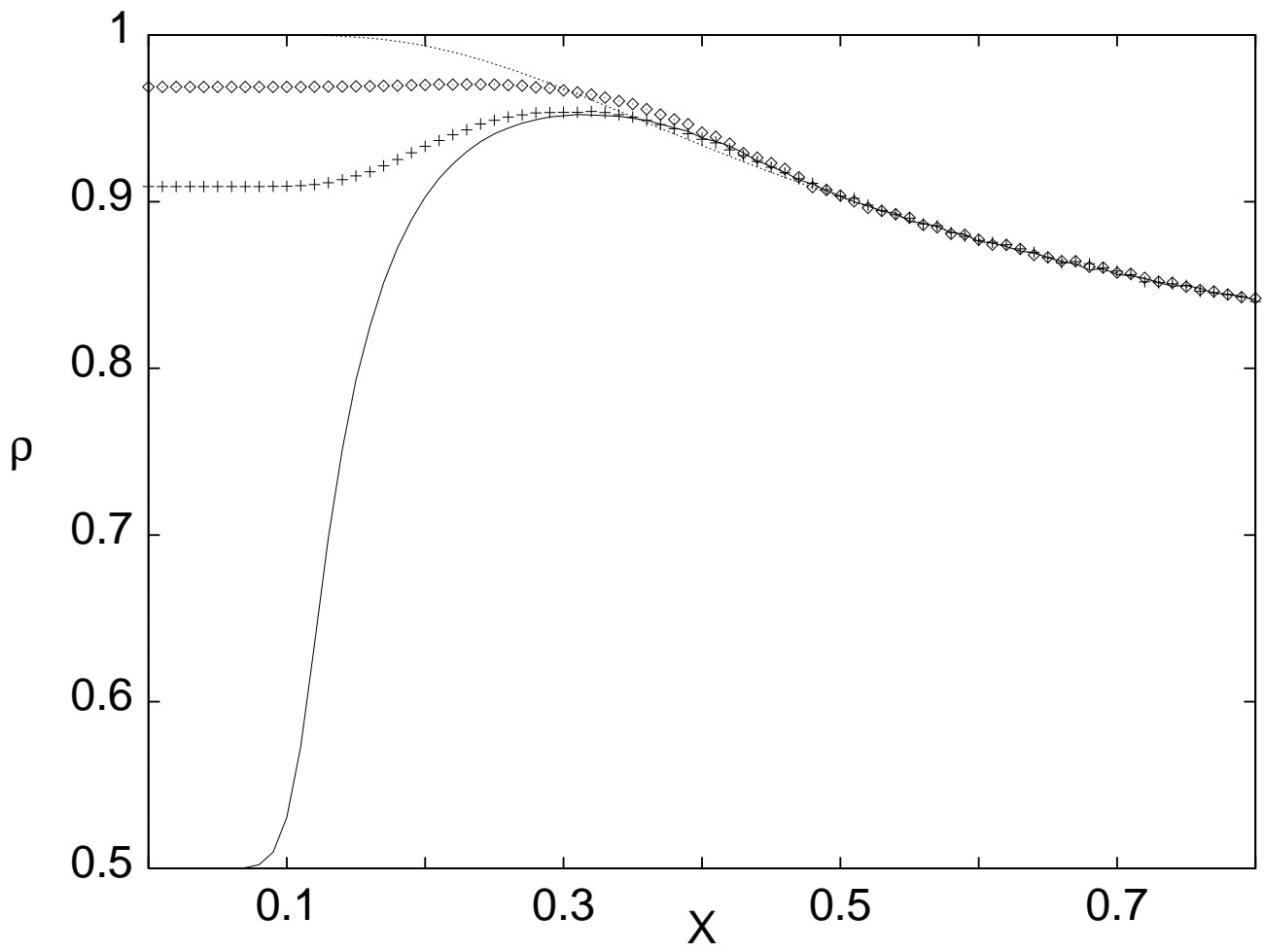
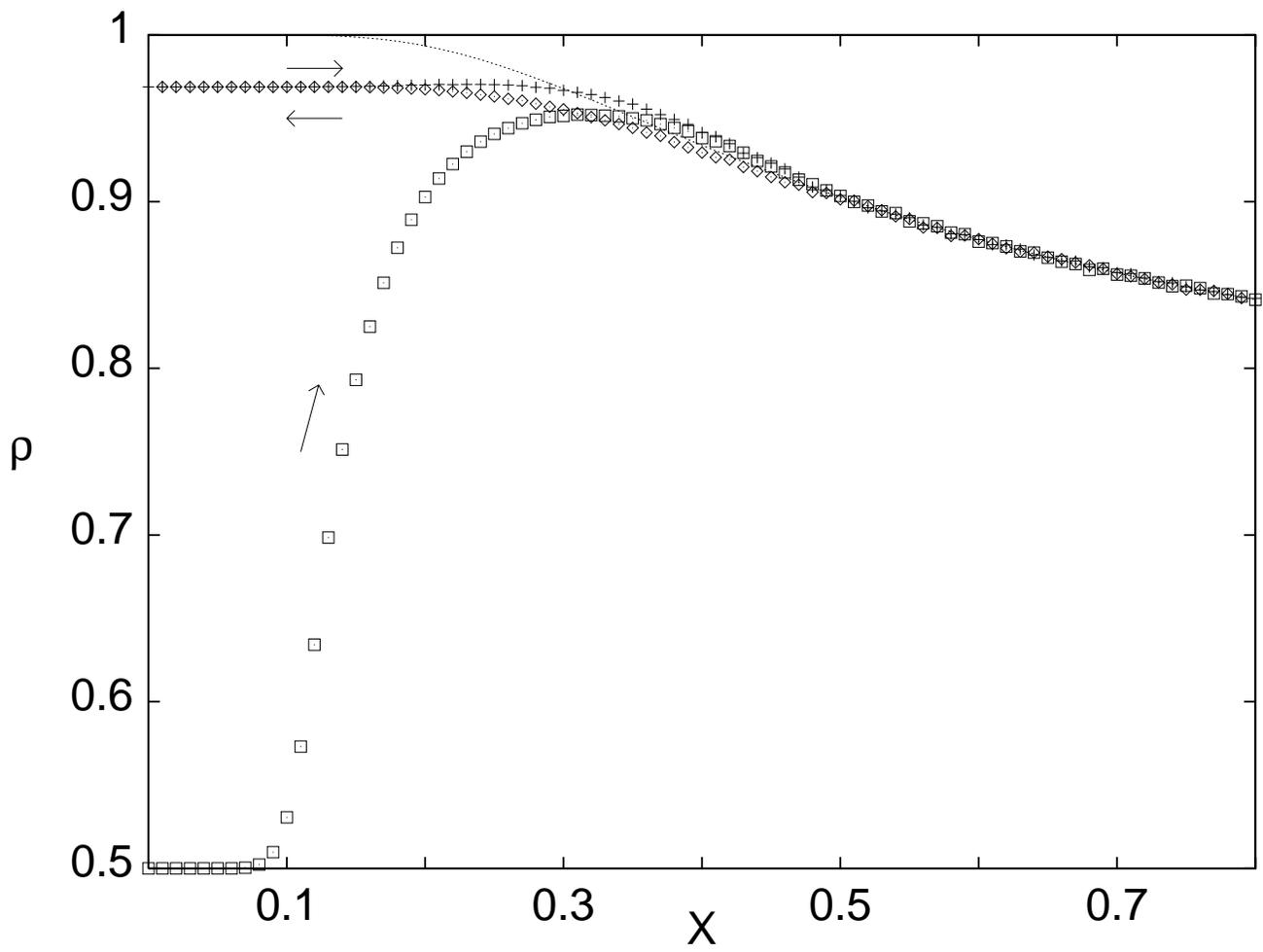
\end{document}